\documentclass[runningheads]{llncs}

\usepackage{eccv}

\usepackage{eccvabbrv}

\usepackage{graphicx}
\usepackage{booktabs}

\usepackage[accsupp]{axessibility}  % Improves PDF readability for those with disabilities.

\usepackage[pagebackref,breaklinks,colorlinks,citecolor=eccvblue]{hyperref}
\usepackage{orcidlink}

\usepackage{tabularx}

\usepackage{xcolor,colortbl}

\begin{document}

\title{Gender Bias Evaluation in Text-to-image Generation: A Survey} 


% Second Author\inst{2,3}\orcidlink{1111-2222-3333-4444} \and
% Third Author\inst{3}\orcidlink{2222--3333-4444-5555}}

\author{Yankun Wu\orcidlink{0009-0005-7175-8307} \and
Yuta Nakashima\orcidlink{0000-0001-8000-3567} \and
Noa Garcia\orcidlink{0000-0002-9200-6359}}

\authorrunning{Wu et al.}

% Springer Heidelberg, Tiergartenstr.~17, 69121 Heidelberg, Germany

\institute{Osaka University \\
\email{\{yankun@is., n-yuta@, noagarcia@\}ids.osaka-u.ac.jp}}

\maketitle

\vspace{-8pt}
\section{Introduction}
\label{sec:intro}
\vspace{-5pt}
The rapid development of text-to-image generation has brought rising ethical considerations, especially regarding gender bias. Given a text prompt as input, text-to-image models generate images according to the prompt. Pioneering models such as Stable Diffusion \cite{rombach2022high} and DALL-E 2 \cite{ramesh2022hierarchical} have demonstrated remarkable capabilities in producing high-fidelity images from natural language prompts. However, these models often exhibit gender bias, as studied by the tendency of generating \textit{man} from prompts such as ``\texttt{a photo of a software developer}'' \cite{luccioni2023stable, cho2023dall, bianchi2023easily}. 
Given the widespread application and increasing accessibility of these models, bias evaluation is crucial for regulating the development of text-to-image generation. Unlike well-established metrics for evaluating image quality or fidelity, the evaluation of bias presents challenges and lacks standard approaches. Although biases related to other factors, such as skin tone, have been explored \cite{bianchi2023easily, bakr2023hrs, naik2023social, katirai2024situating}, gender bias remains the most extensively studied. In this paper, we review recent work on gender bias evaluation in text-to-image generation, involving bias evaluation setup (\cref{sec:setup}), bias evaluation metrics (\cref{sec:metrics}), and findings and trends (\cref{sec:findings}). We primarily focus on the evaluation of recent popular models such as Stable Diffusion, a diffusion model \cite{ho2020denoising} operating in the latent space and using CLIP text embedding \cite{radford2021learning}, and DALL-E 2, a diffusion model leveraging Seq2Seq architectures like BART \cite{lewis2020bart}. By analyzing recent work and discussing trends, we aim to provide insights for future work.

%%%%%%%%%%%%%%%%%%%%%%%%%%%%%%%%%%%%%%%%%%%%%%%%%%%%%%%%%%%%%%%%%%%%%%%%%%%%%%%
\vspace{-6pt}
\section{Bias evaluation setup}
\label{sec:setup}
\vspace{-5pt}
An overview of gender bias evaluation methods in text-to-image generation is shown in \cref{tab:overview}. This section outlines the key components of bias evaluation setups: gender and bias definitions, prompt design, and attribute classification. 

\vspace{-5pt}
\paragraph{Gender definition}
The majority of studies focus on binary gender (\textit{female/woman}, \textit{male/man}) \cite{bakr2023hrs, teo2024measuring, lee2023holistic}. 
Some work employ more than two genders, such as non-binary gender \cite{luccioni2023stable} or neutral gender (\textit{person/people}) \cite{chinchure2023tibet, zhang2023auditing, wu2024stable}.

\vspace{-6pt}
\paragraph{Bias definition}
Derived from definitions in previous work \cite{hirota2023model}, we identify two types of gender bias in text-to-image generation:
1) \textbf{context-to-gender} bias: when gender-neutral prompts do not result in equal probabilities of generating images of woman and man \cite{bakr2023hrs, cho2023dall, wang2024new};
2) \textbf{gender-to-context} bias: when gender-indicating words in the prompt result in significant differences in the context of the generated image (\eg, background, objects) \cite{mannering2023analysing, wu2024stable, zhang2023auditing}.
\begin{table*}[t]
\hspace{-22pt}
\centering
\renewcommand{\arraystretch}{1.1}
\setlength{\tabcolsep}{5pt}
\small
\footnotesize
\caption{Gender bias evaluation studies in text-to-image generation. Note that ``Gender'' in the Attribute column is based on the appearance of people in the generated images .}

\vspace{-5pt}
\scalebox{0.59}{
\begin{tabularx}{1.7\textwidth}{@{}l p{0.21\textwidth} p{0.25\textwidth} p{0.18\textwidth} p{0.25\textwidth} p{0.27\textwidth} @{}}
\toprule
Method & Prompt source & Prompt scenario & Attribute & Attribute extraction & Evaluation \\
\midrule 

Luccioni \etal \cite{luccioni2023stable} & Template & Profession, Identity & Gender & VQA, Image captioning, Clustering & Statistics \\

Bakr \etal \cite{bakr2023hrs}  & LLM & Object & Gender & ArcFace \cite{deng2019arcface}, RetinaFace \cite{deng2020retinaface}, Dex \cite{rothe2015dex} & Mean Absolute Deviation (MAD) \cite{pearson1894contributions} \\

Teo \etal \cite{teo2024measuring} & Template & - & Gender & Classifier & Statistical model \cite{teo2024measuring} \\ 

Lee \etal \cite{lee2023holistic} - Fairness & Dataset & MS-COCO \cite{lin2014microsoft} & Embedding & CLIP \cite{radford2021learning} & CLIPScore \cite{hessel2021clipscore}, Human evaluation\\ 

Lee \etal \cite{lee2023holistic} - Bias & Template & Adjective, Profession & Gender & CLIP \cite{radford2021learning} & L1 distance \\ 

Cho \etal \cite{cho2023dall} & Template & Profession & Gender & BLIP 2 \cite{li2023blip} & MAD \cite{pearson1894contributions} \\ 

Bianchi \etal \cite{bianchi2023easily} & Template & Profession & Gender & CLIP \cite{radford2021learning}  &  Statistics \\

Wang \etal \cite{wang2023t2iat} - Profession & Template & Profession & Gender & CLIP \cite{radford2021learning}  &  Statistics \\

Wang \etal \cite{wang2023t2iat} - Science/Career & Template & Science/Career & Gender & CLIP \cite{radford2021learning}  & T2IAT \cite{wang2023t2iat} \\ 

Chinchure \etal \cite{chinchure2023tibet} & Designed prompt, Dataset, LLM & Creative prompts, Diffusion DB \cite{wang2023diffusiondb} & Concept & VQA & CAS \cite{chinchure2023tibet}, MAD \cite{pearson1894contributions} \\ 

Zhang \etal \cite{zhang2023auditing} & Template & Attire, Activity & Attire & Classifier & GEP \cite{zhang2023auditing} \\ 

Naik \etal \cite{naik2023social} & Template & Adjective, Profession & Gender & Human &  Statistics \\ 

Wu \etal \cite{wu2024stable} & Dataset, LLM & MS-COCO \cite{lin2014microsoft}, GCC \cite{sharma2018conceptual}, Flickr30k \cite{young2014image}, TextCaps \cite{sidorov2020textcaps}  & Embedding, Object & Visual grounding & Bias score \cite{zhao2017men}, Chi-square test, Similarity\\ 

Friedrich \etal \cite{friedrich2024multilingual} & Template & Adjective, Profession, Multilingualism & Gender & Classifier & MAD \cite{pearson1894contributions} \\ 

Sathe \etal \cite{sathe2024unified} & LLM & Profession & Gender & BLIP 2 \cite{li2023blip} & Neutrality \cite{sathe2024unified} \\ 

Luo \etal \cite{luo2024bigbench} & Template & Profession, Social relation, Characteristic & Gender & InternVL \cite{chen2024internvl} & Bias score \cite{luo2024bigbench} \\ 

Chen \etal \cite{chen2024evaluating} & Template & Action, Appearances & Gender & VQA & Statistics \\ 

Wan \etal \cite{wan2024male} & Template & Two professions & Gender & Human & Stereotype Score \cite{wan2024male} \\ 

Wang \etal \cite{wang2024new} & Template & Activity, Object, Adjective, Profession & Gender & Face analyses API & Bias score \cite{wang2024new} \\ 

D{'}Inc{\`a} \etal \cite{d2024openbias} & Dataset & Flickr30k \cite{young2014image}, MS-COCO \cite{lin2014microsoft} & Gender & VQA & Bias Severity Score \cite{d2024openbias} \\ 

Mannering \cite{mannering2023analysing} & Template & Vague scenario & Object & VQA & Chi-square test \\ 

Garcia \etal \cite{garcia2023uncurated} & Dataset & PHASE \cite{garcia2023uncurated} & Safety & Safety checker &  Statistics \\

\bottomrule
\end{tabularx}}
\vspace{-12pt}
\label{tab:overview}
\end{table*}

\vspace{-8pt}
\paragraph{Prompt design}
The prompt, which serves as the semantic guidance of text-to-image generation, is a crucial aspect of bias evaluation. Most of work apply template-based prompts, such as ``\texttt{a photo of [DESCRIPTION]}'', where \texttt{[DESCR-\\IPTION]} may include professions \cite{luccioni2023stable, cho2023dall, lee2023holistic, bianchi2023easily, naik2023social, friedrich2024multilingual, luo2024bigbench, wan2024male, wang2024new}, adjectives \cite{lee2023holistic, naik2023social, friedrich2024multilingual, wang2023t2iat}, or activities \cite{zhang2023auditing, wang2024new}. These variations allow for the investigation of context-to-gender bias across different scenarios. 
Beyond templates, captions in the vision-language dataset (\eg, MS-COCO \cite{lin2014microsoft}) are also employed as natural language prompts \cite{d2024openbias, wu2024stable}.  
Additionally, large language models (LLM) \cite{brown2020language} are increasingly used for prompt generation \cite{bakr2023hrs, chinchure2023tibet, wu2024stable, sathe2024unified}.
Complementing natural language prompts, some studies generate counterfactual prompts by swapping the gender-neutral word (\eg, \textit{person}) with gendered term (\eg, \textit{woman}, \textit{man}) \cite{wu2024stable, cho2023dall, chinchure2023tibet}. This allows for comparison of generated images when only the gender reference is changed.

\vspace{-5pt}
\paragraph{Attribute classification}
To evaluate gender bias in the generated images, protected attributes like gender need to be assigned for the representations of people in the generated images based on their appearance.
To assign gender to generated images, one approach involves applying a gender classifier to generated faces \cite{bakr2023hrs, friedrich2024multilingual}. 
Other work \cite{lee2023holistic, cho2023dall, bianchi2023easily, wang2023t2iat, sathe2024unified, luo2024bigbench} extend beyond facial analysis by utilizing embeddings of the entire image from vision-language models (\eg CLIP \cite{radford2021learning}, BLIP 2 \cite{li2023blip}, InternVL \cite{chen2024internvl}, etc). These embeddings are compared to the text embeddings of sentences like ``\texttt{a photo of a woman/man},'' and the gender of the highest similarity sentence is assigned to the image \cite{cho2023dall, zhang2023iti, bianchi2023easily}.
Another popular method for obtaining gender is using visual question answering (VQA) models, by asking questions like ``\texttt{What is the gender of the person?}'' \cite{chen2023minigpt}. 
Additionally, other work rely on human annotation for assigning gender to generated images \cite{wan2024male, naik2023social}. 
Beyond protected attributes, some studies examine other attributes such as attire \cite{zhang2023auditing, cho2023dall}, concepts \cite{chinchure2023tibet}, and objects \cite{mannering2023analysing, wu2024stable}. Garcia \etal \cite{garcia2023uncurated} generate images from the captions and obtain their safety label by the Safety Checker module in Stable Diffusion.

In conclusion, evaluating text-to-image generation requires careful design tailored to specific objectives. Future work should set clear definitions of gender and bias, design appropriate prompts, and identify the attributes to be examined.

%%%%%%%%%%%%%%%%%%%%%%%%%%%%%%%%%%%%%%%%%%%%%%%%%%%%%%%%%%%%%%%%%%%%%%%%%%%%%%%
\vspace{-5pt}
\section{Bias evaluation metrics}
\label{sec:metrics}
\vspace{-5pt}
After obtaining the protected attributes, evaluation metrics are employed according to specific objectives. We categorize these metrics into three types: distribution metrics, bias tendency metrics, and quality metrics. Distribution metrics address context-to-gender bias by statistically analyzing differences in distributions. Bias tendency metrics correspond to gender-to-context bias, examining whether the attributes such as objects or concepts exhibit bias toward a specific gender. Quality metrics include standard text-to-image generation metrics, such as semantic alignment and image quality.

\vspace{-5pt}
\paragraph{Distribution metrics}

To measure differences between the detected attributes distribution and an unbiased distribution, several studies apply Mean Absolute Deviation (MAD) \cite{pearson1894contributions} to attributes like perceived gender \cite{bakr2023hrs, cho2023dall, friedrich2024multilingual}. 
Additionally, some work use the chi-square test to examine to determine whether there are significant differences between objects generated from counterfactual prompts where only the gender indicator varies \cite{mannering2023analysing, wu2024stable}. 

\vspace{-5pt}
\paragraph{Bias tendency metrics}
To determine if the detected attributes are biased toward a certain gender, the metrics used vary depending on the specific attributes.
For protected attributes like gender, several approaches assess whether a certain gender is prone to be generated from gender-neutral prompts. One common method involves calculating the proportion of predicted genders relative to the total number of samples \cite{bianchi2023easily, wang2023t2iat, naik2023social, luccioni2023stable}. 
Additionally, some studies compare these proportions with real-world data, such as from the U.S. Bureau of Labor Statistics, to determine if the model amplifies bias in the real world \cite{bianchi2023easily, wang2023t2iat}. 
Luo \etal \cite{luo2024bigbench} compute the cosine similarity between the proportions of gender in the generated images and in the prompts. 
Wang \etal \cite{wang2024new} propose a bias score based on gender in generated images relative to input images.
The Neutrality metrics, proposed by Sathe \etal \cite{sathe2024unified}, introduces ``no preference'' class alongside binary genders. This metric assesses whether the model exhibits bias toward generating a specific gender from neutral prompts. 
Wan \etal \cite{wan2024male} use a template involving two gender-stereotypical professions to develop Stereotype Score, which evaluates whether the gender in the generated images aligns with these stereotypes. 
To investigate object-related differences and concept evaluation regarding gender, Zhang \etal \cite{zhang2023auditing} propose Gender Presentation Differences (GEP) metric. Wu \etal \cite{wu2024stable} use Bias score \cite{zhao2017men} and co-occurrence similarity metrics to analyze object co-occurrence differences between genders.
Wang \etal \cite{wang2023t2iat} employ WEAT-like metric \cite{caliskan2017semantics} to determine if two attribute sets (\eg science and art) have significant differences in their associations with gender.
In the context of safety, Garcia \etal \cite{garcia2023uncurated} use the gender annotation of original images to analyze the proportion of generated images that are labeled as unsafe.

\vspace{-5pt}
\paragraph{Quality metrics}
\vspace{-5pt}
In addition to bias-related metrics, standard metrics for text-to-image generation are also employed. For example, Lee \etal \cite{lee2023holistic} apply CLIPScore \cite{hessel2021clipscore} and human evaluation \cite{otani2023toward} to assess the semantic alignment between images and prompts, focusing on gender differences.
Naik \etal \cite{naik2023social} employ Fréchet Inception Distance (FID) \cite{heusel2017gans} to compare generated images with real images under gendered prompts related to professions.

%%%%%%%%%%%%%%%%%%%%%%%%%%%%%%%%%%%%%%%%%%%%%%%%%%%%%%%%%%%%%%%%%%%%%%%%%%%%%%%
\vspace{-6pt}
\section{Findings and trends}
\vspace{-6pt}
\label{sec:findings}
The most common models to be evaluated are Stable Diffusion \cite{rombach2022high} and DALL-E 2 \cite{ramesh2022hierarchical}. While most studies focus on two or three models, Lee \etal \cite{lee2023holistic} evaluate $26$ models, and Luo \etal \cite{luo2024bigbench} evaluate $11$ models. 
A consistent finding across various studies is that models tend to generate \textit{man} more frequently for professions. This tendency is observed in Stable Diffusion \cite{luccioni2023stable, cho2023dall, wang2023t2iat, bianchi2023easily, chinchure2023tibet}, DALL-E \cite{luccioni2023stable}, minDALL-E \cite{kakaobrain2021minDALL-E, cho2023dall}, and Karlo \cite{kakaobrain2022karlo-v1-alpha, cho2023dall}. Furthermore, specific professions, such as singers and authors, may exhibit different bias tendencies \cite{cho2023dall, naik2023social}.
Beyond gender bias in professions, models like Stable Diffusion, Karlo, and minDALL-E have shown a tendency to generate gender-specific attire, such as skirts for \textit{woman} and suits for \textit{man} \cite{zhang2023auditing, wu2024stable, cho2023dall, d2024openbias}. 
When examining gender generated from neutral prompts, Bakr \etal \cite{bakr2023hrs} observe slight biases in DALL-E 2, Cogview 2 \cite{ding2022cogview2}, and minDALL-E. Conversely, Lee \etal \cite{lee2023holistic} found that minDALL-E, DALL-E mini \cite{Dayma_DALL·E_Mini_2021}, and SafeStableDiffusion \cite{schramowski2023safe} exhibited the least bias, while Dreamlike Diffusion \cite{dreamlike}, DALL-E 2 \cite{ramesh2022hierarchical}, and Redshift Diffusion \cite{redshift} showed more severe biases.
Moreover, some studies report that bias is not limited to regions containing humans but also extends to the overall image context, including objects and background elements \cite{wu2024stable, d2024openbias, mannering2023analysing}.

An emerging trend is the increasing comprehensiveness of model evaluations, with a broader range of models, diverse prompts, and multiple axes of bias assessment \cite{bakr2023hrs, luo2024bigbench, lee2023holistic}. Recent work is also focusing on a more detailed examination of bias sources, offering valuable insights for future bias mitigation methods \cite{wu2024stable}.

%%%%%%%%%%%%%%%%%%%%%%%%%%%%%%%%%%%%%%%%%%%%%%%%%%%%%%%%%%%%%%%%%%%%%%%%%%%%%%%
\vspace{-6pt}
\section{Conclusion}
\vspace{-5pt}
In this paper, we reviewed recent research on gender bias in text-to-image generation, focusing on evaluation setup (including definitions of gender and bias, prompts, and attributes) and evaluation metrics. We summarised key findings and analyzed emerging trends. We hope this survey provides insights for future work in text-to-image generation.

% \clearpage\mbox{}Page \thepage\ of the manuscript.
% \clearpage\mbox{}Page \thepage\ of the manuscript.
% \clearpage\mbox{}Page \thepage\ of the manuscript.
% \clearpage\mbox{}Page \thepage\ of the manuscript.
% \clearpage\mbox{}Page \thepage\ of the manuscript. This is the last page.
% Now we have reached the maximum length of an ECCV \ECCVyear{} submission (excluding references).
% References should start immediately after the main text, but can continue past p.\ 14 if needed.
% \clearpage  % TODO REVIEW/FINAL: This \clearpage needs to be removed from both review and camera-ready versions.

\bibliographystyle{splncs04}
\bibliography{main}
\end{document}